# Crystallite Size Effect on Lattice Strain and Crystal Structure of $Ba_{1/4}Sr_{3/4}MnO_3$ Layered Perovskite Manganite

Dinesh Kumar[a)], Monika Singh[b)] and Akhilesh Kumar Singh[c)*]

*School of Materials Science and Technology, Indian Institute of Technology (Banaras Hindu University), Varanasi-221005, India*
Email: [a)]dineshiitbhu@gmail.com, [b)]monikas.rs.mst15@itbhu.ac.in, [c)*]aksingh.mst@iitbhu.ac.in

**Abstract.** The single phase polycrystalline $Ba_{1/4}Sr_{3/4}MnO_3$ layered perovskite manganite has been synthesized by combustion method having various crystallite sizes. The room temperature X-ray diffraction patterns reveal that $Ba_{1/4}Sr_{3/4}MnO_3$ crystallizes into hexagonal crystal structure with space group $P6_3/mmc$ as confirmed by Rietveld refinement. The scanning electron micrographs of $Ba_{1/4}Sr_{3/4}MnO_3$ reveal uniform crystallite size of the samples. Effect of crystallite size on lattice strain and crystal structure has been studied using Rietveld refinement and Williamson-Hall plot, respectively. The structural lattice parameters decrease with increasing crystallite size. However, lattice strain increases with increasing crystallite-size.

"**Keywords:** Perovskite; Manganite; X-ray diffraction; Rietveld refinement; Williamson-Hall plot Method."

## INTRODUCTION

In the past few years, significant attempts have been dedicated to investigate structural, magnetic and electrical properties of nanocrystalline rare earth perovskite manganites and determining how these properties depend on crystallite size [1-3]. The rare earth layered manganites crystallize into distorted hexagonal or orthorhombic crystal structures and can exhibit both ferroelectric and magnetic ordering as well as magnetoelectric coupling [3]. Most of the hexagonally distorted layered manganites exhibit ferroelectric transitions at high temperatures ($T_C$ = 600-1000 K) and antiferromagnetic (AFM) transitions at low temperatures ($T_N$ = 70-130 K) [4]. The layered perovskite manganites having chemical formula $AMnO_3$, where A is divalent alkaline or trivalent rare earth cation(s) (A = Ba, Sr, Ca, La, Nd, Pr), show antiferromagnetic ordering due to superexchange interactions between $Mn^{4+}$ or $Mn^{3+}$ ions via $O^{2-}$ ions i.e. $Mn^{4+}$ -$O^{2-}$ -$Mn^{4+}$ or $Mn^{3+}$ -$O^{2-}$ -$Mn^{3+}$[5-6].

Recently, Shankar et al. [1] studied crystallite size-dependent structural properties of various half doped manganites and reported that unit cell lattice parameters increase with decreasing crystallite size. However, Kumar et al. [7] reported that the lattice parameters of $Pr_{0.6}Ca_{0.4}MnO_3$ manganite decrease with the decrease of crystallite size. The probable structural phase and phase transformations in the perovskite $ABO_3$ can be predicted using structural tolerance factor t, (t = $d_{A-O}/\sqrt{2}\, d_{B-O}$) given by Goldschmidt [8]. There is very little literature available on $Ba_{1-x}Sr_xMnO_3$ perovskite manganite, so, for this reason, we have chosen this system for investigating structure and properties in nanocrystalline samples [9,11]. The structural unit cell lattice parameters for $Ba_{1-x}Sr_xMnO_3$ is expected to lie between the lattice parameters of end components $BaMnO_3$ and $SrMnO_3$ manganites [10].

Here, in the present article, we have studied effect of crystallite size on structural lattice parameters and lattice strain of $Ba_{1/4}Sr_{3/4}MnO_3$ layered perovskite manganite synthesized by combustion method. The structural parameters (unit cell constants, unit cell volume, bond-lengths and tolerance factor) were determined by Rietveld structure refinement using powder x-ray diffraction data. Williamson-Hall plot method has been used to calculate the crystallite size and lattice strain.

## EXPERIMENTAL DETAILS

The polycrystalline $Ba_{1/4}Sr_{3/4}MnO_3$ layered perovskites were synthesized via combustion method [10]. In this article, $Ba_{1/4}Sr_{3/4}MnO_3$ is denoted by BSMO. The stoichiometric amount of $BaCO_3$ (99.0%, Sigma Aldrich) and $SrCO_3$ (98.0%, Sigma Aldrich) were dissolved in $HNO_3$ (69.0%, Himedia) to form their nitrates according to the chemical reactions (1) and (2):

$$BaCO_3 + 2HNO_3 \rightarrow Ba(NO_3)_2 + CO_2 + H_2O \ldots (1)$$
$$SrCO_3 + 2HNO_3 \rightarrow Sr(NO_3)_2 + CO_2 + H_2O \ldots (2)$$

While, manganese acetate tetrahydrate (99.0%, Sigma Aldrich) and glycine (99.5%, Himedia) used as fuel, were dissolved in distilled water. All precursor solutions were mixed in a large beaker and heated on the magnetic stirrer hot plate at 200°C under continuous stirring. After few hours of constant stirring, auto-ignition occurs resulting black-brown powder. In the presence of glycine the final chemical reaction takes place according to the reaction given by equation (3):

$$¼ Ba(NO_3)_2 + ¾ Sr(NO_3)_2 + (CH_3COO)_2Mn.4H_2O \rightarrow Ba_{1/4}Sr_{3/4}MnO_3 + 3CO_2 + CO + 2NH_3 + 4H_2O \ldots (3)$$

The resultant black-brown powder was calcined at various temperatures in the temperature range 900-1200°C for 6h, to obtain nanocrystalline powder samples with different crystallite sizes. For the analysis of the structure of the samples, powder X-ray diffraction (XRD) patterns were recorded using X-ray diffractometer (Rigaku, Miniflex600) in the 2θ range of 20°-120° @ 2°/min at steps of 0.02°. Scanning electron microscopy (SEM) and energy dispersive X-ray spectroscopy (EDS) were performed for microstructural and elemental analysis, respectively. The XRD patterns were analysed by Rietveld structure refinement method using FullProf Suite [11].

## RESULTS AND DISCUSSION

### Structural Refinement: Rietveld Analysis

The room temperature XRD patterns of BSMO layered perovskite samples calcined at various temperatures (S1@900°C, S2@950°C, S3@1000°C, S4@1050°C, S5@1100°C & S6@1200°C) are plotted in Figure 1(a). Very Small impurities peaks of BaO and SrO were detected near 29.7° and 32.3°, respectively, in the XRD patterns of the samples S2-S5 due to insolubility of BaO and SrO. Remaining XRD peaks for all the patterns can be indexed using hexagonal crystal structure with space group $P6_3/mmc$ [10]. The inset of Figure 1(a) shows selected Bragg reflections (110) between 2θ angle 32.0° and 33.5° for the samples S1 and S6. The Bragg peak (110) get broadened with decreasing calcination temperature from 1200°C to 900°C, showing decrease in crystallite size and shifted towards lower angle side due to expansion in lattice parameters with decrease of calcination temperature. Variation of lattice parameters (lattice constants and unit cell volume) as a function of calcination temperature (crystallite size) is shown in Figure 1(b).

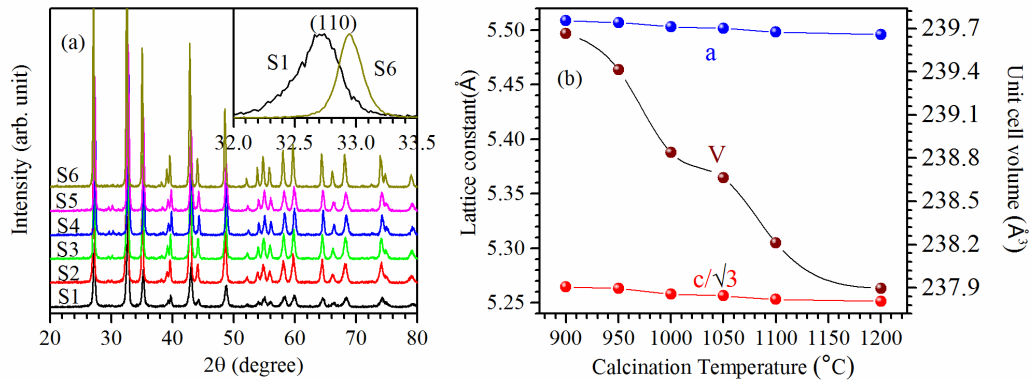

**FIGURE 1**: (a) Room temperature XRD patterns for BSMO samples calcined at various temperatures. Inset shows selected Bragg peaks for the samples S1 and S6. (b) Variation of lattice parameters and unit cell volume as a function of calcination temperature.

Rietveld structural analysis of BSMO for different crystallite sizes was performed by using XRD data. In the process of Rietveld refinement, we used Pseudo-Voigt profile function to model the XRD peak profiles, whereas, the background was modelled using six coefficient polynomial. In the hexagonal crystal structure with space group $P6_3/mmc$, we considered $Ba^{2+}(1)/Sr^{2+}(1)$ ions at site 2(a) (0, 0, 0), $Ba^{2+}(2)/Sr^{2+}(2)$ ions at site 2(c) (1/3, 2/3, 1/4), $Mn^{4+}$ ions at sites 4(f) (1/3, 2/3, 1/2+$\delta z$), $O^{2-}(1)$ ions at sites 6(g) (1/2, 0, 0) and $O^{2-}(2)$ ions at sites 6(h) (-$\delta x$, -2$\delta x$, 1/4) [10]. The Rietveld fit between observed and calculated patterns shows very good fit with agreement parameters $R_f$ = 5.28 % and $R_B$ = 6.79 % is shown in Figure 2(a) for the sample S3. Dots in Figure 2(a) show observed pattern, while the calculated pattern is shown by continuous line. The lower curve indicates difference between observed and calculated XRD patterns. Vertical bars indicate positions of Bragg's peaks. Figure 2(b) shows Ball and Stick model of the unit cell which shows alternate layers of "Mn" and "Ba-O/Sr-O", for the BSMO layered perovskite. The inset of Figure 2(a) shows the SEM micrograph for polycrystalline samples S6. The average grain size was calculated and found to be ~1μm.

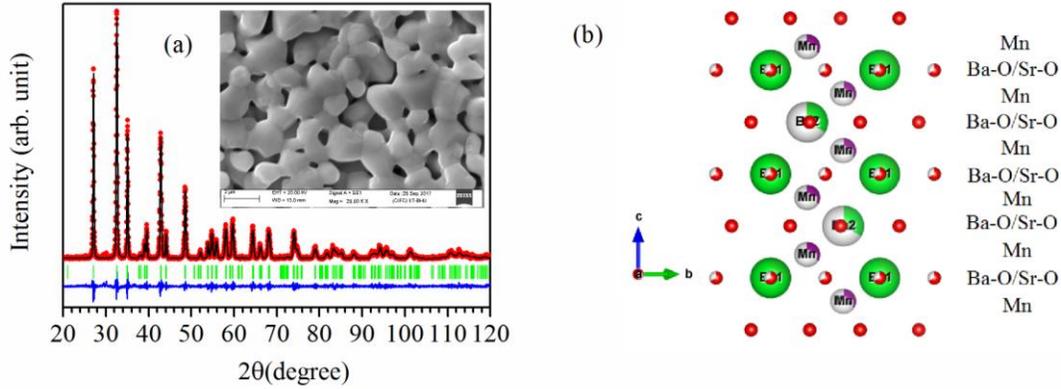

**FIGURE 2**: (a) Rietveld fit for the polycrystalline sample of BSMO calcined at 1000°C. The inset shows SEM micrograph for the sample S6. (b) Ball and Stick model for BSMO layered perovskite manganite.

The refined lattice constant "a" varies from a = 5.5090(2) Å for S1 to a = 5.4957(2) Å for S6 and lattice constant "c" varies from c = 9.1188(3) Å for S1 to c = 9.0950(3) Å for S6. The unit cell volume decreases from V = 239.66(1) Å$^3$ for S1 to V = 237.89(1) Å$^3$ for S6. The average bond length <$d_{Ba-O}$> decreases from 2.7889 Å to 2.7718 Å, while the average bond length <$d_{Mn-O}$> increases from 1.8800 Å to 1.9144 Å on increase of crystallite size from 22 nm (sample S1) to 212 nm (sample S6), respectively. The calculated value of structural tolerance factor decreases as a function of crystallite size (calcination temperature) from 1.049 for sample S1 to 1.024 for sample S6. The details of the structural parameters for the synthesized samples are given in Table 1.

**TABLE 1**. Structural parameters, strain and tolerance factor for $Ba_{1/4}Sr_{3/4}MnO_3$ compound obtained after Rietveld structure refinement using XRD data.

| Sample | d (nm) | a (Å) | c (Å) | V (Å$^3$) | ε | $\chi^2$ | t |
|---|---|---|---|---|---|---|---|
| S1 | 22 | 5.5090(2) | 9.1188(3) | 239.66(1) | 1.32×10$^{-3}$ | 1.85 | 1.049 |
| S2 | 32 | 5.5070(2) | 9.1156(4) | 238.00(2) | 2.07×10$^{-3}$ | 1.64 | 1.022 |
| S3 | 103 | 5.5031(2) | 9.1066(4) | 238.84(2) | 2.37×10$^{-3}$ | 1.39 | 1.023 |
| S4 | 127 | 5.5018(2) | 9.1042(3) | 238.66(1) | 2.74×10$^{-3}$ | 1.57 | 1.023 |
| S5 | 196 | 5.4983(2) | 9.0987(4) | 238.22(2) | 3.23×10$^{-3}$ | 1.51 | 1.022 |
| S6 | 212 | 5.4957(2) | 9.0950(3) | 237.89(1) | 4.58×10$^{-3}$ | 1.53 | 1.024 |

For the calculation of the crystallite sizes and strain of the synthesized samples, we used the Williamson-Hall (W-H) plot method [12] given by the relation (4):
$$\beta \cos\theta = 0.89\ \lambda/d + 4\varepsilon \sin\theta \quad \ldots(4)$$

where, λ is the wavelength of the X-ray radiation used (λ = 1.5406Å), θ is Bragg angle, d is crystallite size, β is the full width at half-maximum (FWHM) of the Bragg peak and ε is the lattice strain. A representative W-H plot for the sample S3 with crystallite size 103 nm is depicted in Figure 3(a).The crystallite sizes calculated using W-H plot method were found to be 22, 32, 103, 127, 195 and 212 nm for samples S1, S2, S3, S4, S5 and S6, respectively, which implies that crystallite size increases with increasing calcination temperature. The value of lattice strain strongly depends on crystallite size and increases with increasing crystallite size as shown in Figure 3(b).

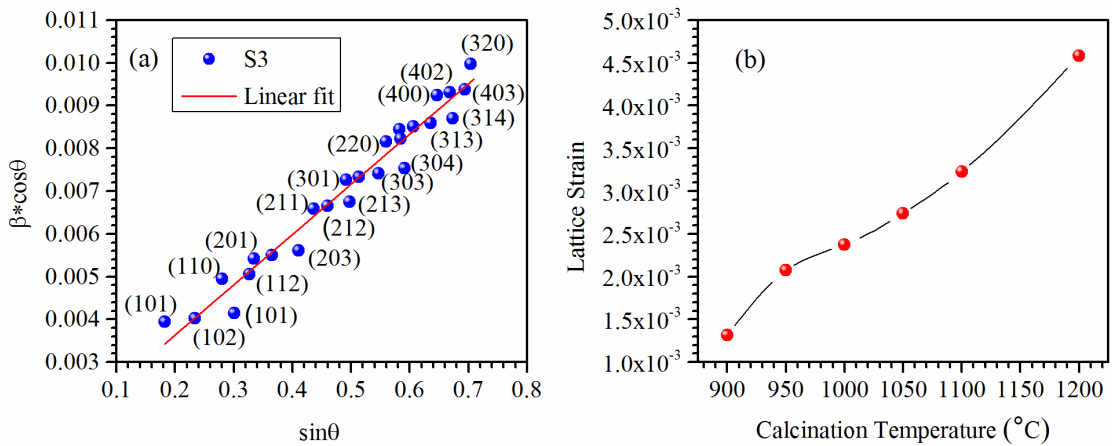

**FIGURE 3.** (a) Williamson Hall plot for the sample S3. (b) Variation of lattice strain as a function of calcination temperature.

## CONCLUSIONS

The nanocrystalline $Ba_{1/4}Sr_{3/4}MnO_3$ layered perovskite manganite has been synthesized successfully using combustion method followed by calcination process to acquire different crystallite sizes. The Rietveld structure refinement using XRD data reveals that BSMO crystallizes in the hexagonal crystal structure with space group $P6_3/mmc$. The lattice constant "a" increases from a = 5.5090(2) Å for sample S1 to a = 5.4957(2) Å for sample S6, while "c" varies from c = 9.1188(3) Å for sample S1 to c = 9.0950(3) Å for sample S6. The value of unit cell volume decreases from V = 239.66(1) Å$^3$ for S1 to V = 237.89(1) Å$^3$ for S6. The value of lattice strain varies from $1.32 \times 10^{-3}$ for sample S1 to $4.58 \times 10^{-3}$ for sample S6, while structural tolerance factor decreases from 1.049 to 1.024.